\documentclass[12pt,a4paper]{article}
\usepackage[T1]{fontenc}
\usepackage[utf8]{inputenc}
\usepackage{booktabs}
\usepackage{amsmath}
\usepackage{graphicx}
\usepackage{hyperref}
\usepackage{xcolor}
\usepackage{siunitx}
\usepackage{gensymb}

\begin{document}

\title{Adaptive time series analysis of Mauna Loa CO$_2$ data: tvf-EMD based detrend and extraction of seasonal variability}

\author{Stefano Bianchi$^a$, Alessandro Longo$^{a,b}$, Wolfango Plastino$^{a,b}$}

\date{\small \it{$^a$Department of Mathematics and Physics, Roma Tre University, Via della Vasca Navale, 00146, Rome, Italy}\\
\it{$^b$INFN, Sezione di Roma Tre, Via della Vasca Navale 84, 00146 Rome, Italy.}\\
}

\maketitle

\begin{abstract}
Adaptive time series analysis has been applied to investigate variability of CO$_2$ concentration data, sampled weekly at Mauna Loa monitoring station. Due to its ability to mitigate mode mixing, the recent time varying filter Empirical Mode Decomposition (tvf-EMD) methodology is employed to extract local narrowband oscillatory modes. In order to perform data analysis, we developed a Python implementation of the tvf-EMD algorithm, referred to as \emph{pytvfemd}. The algorithm allowed to extract the trend and both the six month and the one year periodicities, without mode mixing, even though the analysed data are noisy. Furthermore, subtracting such modes the residuals are obtained, which are found to be described by a normal distribution. Outliers occurrence was also investigated and it is found that they occur in higher number toward the end of the dataset, corresponding to solar cycles characterised by smaller sunspot numbers. A more pronounced oscillation of the residuals is also observed in this regard, in relation to the solar cycles activity too. 
\end{abstract}

%%%%%%%%%%%%%%%%%%%%%%%%%%%%%%%%%%%%%%%%%%%%%%%%
\section{Introduction}
Carbon dioxide ($CO_2$) is a greenhouse gas which has an impact on the radiative forcing and hence on the planetary radiation budget. Its concentration in the troposphere increased since the pre-industrial era mainly due to the increase in the burning of fossil fuels and to the change of land use, which leads to a diminished capacity by plants to uptake and remove $CO_2$ \cite{Zeng_2005}. The variability of $CO_2$ atmospheric concentration is both dependent on anthropogenic emissions, which contribute to the increase in the average concentration, i.e. in the trend, and on climatic conditions affecting the sources and sinks of this greenhouse gas. While interannual variability is mainly related to changes in climate occurring at large scale \cite{Rayner_1999}, short term variability is related to atmospheric transport processes interacting with regional sources and sinks of $CO_2$ \cite{Artuso_2009}. Precisely quantifying $CO_2$ variability, among with the properties of sources and sinks, has relevant implications in modelling and forecasting \cite{Agusti_2014}. This paper focus on the seasonal variability of $CO_2$ atmospheric concentration data, sampled at Mauna Loa (Hawaii), extracting and quantifying the one year and six month oscillatory modes embedded in the time series, while the trend term is also extracted. The $CO_2$ monitoring station located in Mauna Loa (19.54\degree N, 155.58\degree W) is recording $CO_2$ data at an height of 3397m above sea level since 1958 \cite{Keeling_1978}. The Mauna Loa station is ideal for climate change studies involving greenhouse gases, being both in a remote location and also being slightly affected by the presence of human activity and vegetation. In order to charcterise data variability, an algorithm for adaptive time series analysis referred to as time varying filter EMD (tvf-EMD) was employed \cite{Li_2017}. The tvf-EMD algorithm allows to extract oscillatory modes from time series which are possibly both nonlinear and nonstationary. It has been already applied to characterise atmospheric $^{7}Be$ time series data, sampled by the International Monitoring System of the Comprehensive Nuclear Test Ban Treaty Organisation, in order to extract the yearly oscillatory modes in the data and relate them to the seasonal dynamic of the convective Hadley cell \cite{Longo_2019}, to investigate and denoise seismometer data and to identify culprits of scattered light noise at the Virgo interferometer \cite{Longo_2020, Longo_2020SC}. %For the purpose of this analysis, a Python based version of the tvf-EMD algorithm has been developed (pytvfemd).

%%%%%%%%%%%%%%%%%%%%%%%%%%%%%%%%%%%%%%%%%%%%%%%%
%\clearpage

\section{Methodology}\label{methodology}
The data analysed were collected weekly at Mauna Loa monitoring station. They are $CO_2$ time series data spanning from 1975 to 2015, and were processed using a Python implementation of the tvf-EMD adaptive algorithm (pytvfemd), developed for this purpose. The main features of the adaptive algorithm are hereafter described. Since Hilbert spectral analysis (HSA) is meaningfully applied to nearly monocomponent time series, adaptive algorithms aim at extracting monocomponent oscillatory modes from the data, a good approximation of a monocomponent mode being the so called intrinsic mode functions (IMFs) \cite{Peng_2005}. To be an IMF, an oscillatory mode must fulfill two requirements: it should have zero mean and the number of its extrema must equal the number of its zero crossings. These features ensure that, in most cases, HSA of the modes provides a meaningful estimation of their instantaneous frequency and amplitude. Performing HSA on all the extracted modes the Hilbert-Huang transform is obtained, a time frequency representation of the data which has very high resolution. One of the most widely used algorithms to extract IMFs is the so called Empirical Mode Decomposition (EMD) \cite{Huang_1998}. The EMD procedure to extract IMFs is the following: first the extrema in the data are identified and the upper and lower envelopes built using cubic spline approximation. Then the envelope mean is subtracted from the data. This is iterated until the two aforementioned requirements are met, which is known as sifting, and a first IMF obtained. This IMF is subtracted from the data and the algorithm is repeated. A set of IMF is obtained plus a residual term which represent either the local average or, if present, the trend in the data. The EMD algorithm is complete, i.e. summing the IMFs and the trend term exactly recover the original data. Hence, using EMD a time series can be represented by the following expansion
\begin{equation}
x(t)=\sum_{i=1}^{K}c_{i}(t)+T(t)
\label{EMD}
\end{equation}
where $t=1\dots N$ with $N$ being the length of the data and $c_{i}(t)$ and $T(t)$ being the ith IMF and the trend term, respectively. It is found empirically that the number of extracted IMFs is $K\simeq \log_{2}N$. A drawback known as mode mixing can arise when applying EMD to noisy data \cite{Huang_1998}. It is due to the random distribution of extrema in noisy data, which can possibly affect the envelope computation step of EMD. Mode mixing is present when an IMF contains widely different scales of oscillation or when oscillations of similar scales are contained in more than one IMF. In order to mitigate this problem, the so called time varying filter EMD (tvf-EMD) algorithm was developed \cite{Li_2017}. In tvf-EMD the sifting iterations are stopped when the Loughlin instantaneous bandwidth \cite{Loughlin,Boash2}, which for a two component signal is given by 
\begin{equation}
LIB(t)=\sqrt{\frac{a_{1}'^{2}(t)+a_{2}'^{2}(t)}{a_{1}^{2}(t)+a_{2}^{2}(t)}
+\frac{a_{1}^{2}(t)a_{2}^{2}(t)(\phi'_{1}(t)-\phi'_{2}(t))^{2}}{(a_{1}^{2}(t)+a_{2}^{2}(t))^{2}}}
\label{IB_LOUGH}
\end{equation}
in term of instantaneous frequency and amplitude, of the oscillatory mode becomes smaller than a selected threshold. To this aim, a bandwidth threshold parameter (BWR) is used. This, together with a cutoff frequency realignment algorithm (see Algorithm 2 in \cite{Li_2017}), allows to obtain more narrowband oscillatory modes, effectively mitigating mode mixing.
The tvf-EMD algorithm employs B-splines \cite{Unser_1999} as a time varying filter, another parameter of the algorithm being the spline order $n$, related to the filter frequency roll off.
Regarding the tvf-EMD algorithm, the parameters used for the analysis of the $CO_2$ time series are the following: bandwidth threshold ratio $BWR=0.1$, B-spline order $n=26$. The maximum number of modes to be obtained was set to $K=\log_{2}N$, which is the number of modes typically obtained with standard EMD \cite{Huang_1998}.
%%%%%%%%%%%%%%%%%%%%%%%%%%%%%%%%%%%%%%%%%%%%%%%%
%\clearpage	

\begin{figure}
\centering
\includegraphics[scale=0.38]{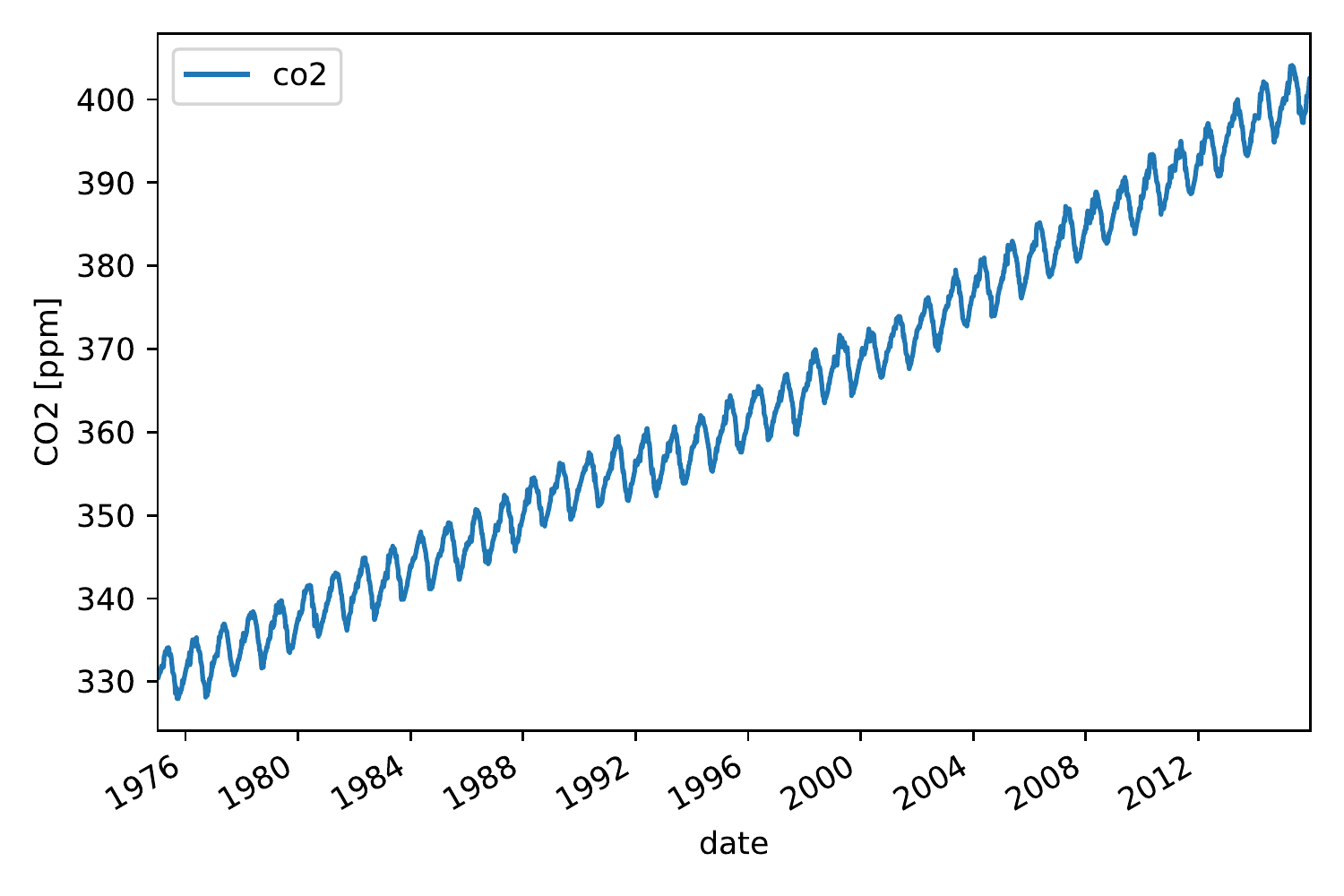}%
\qquad\qquad
\includegraphics[scale=0.38]{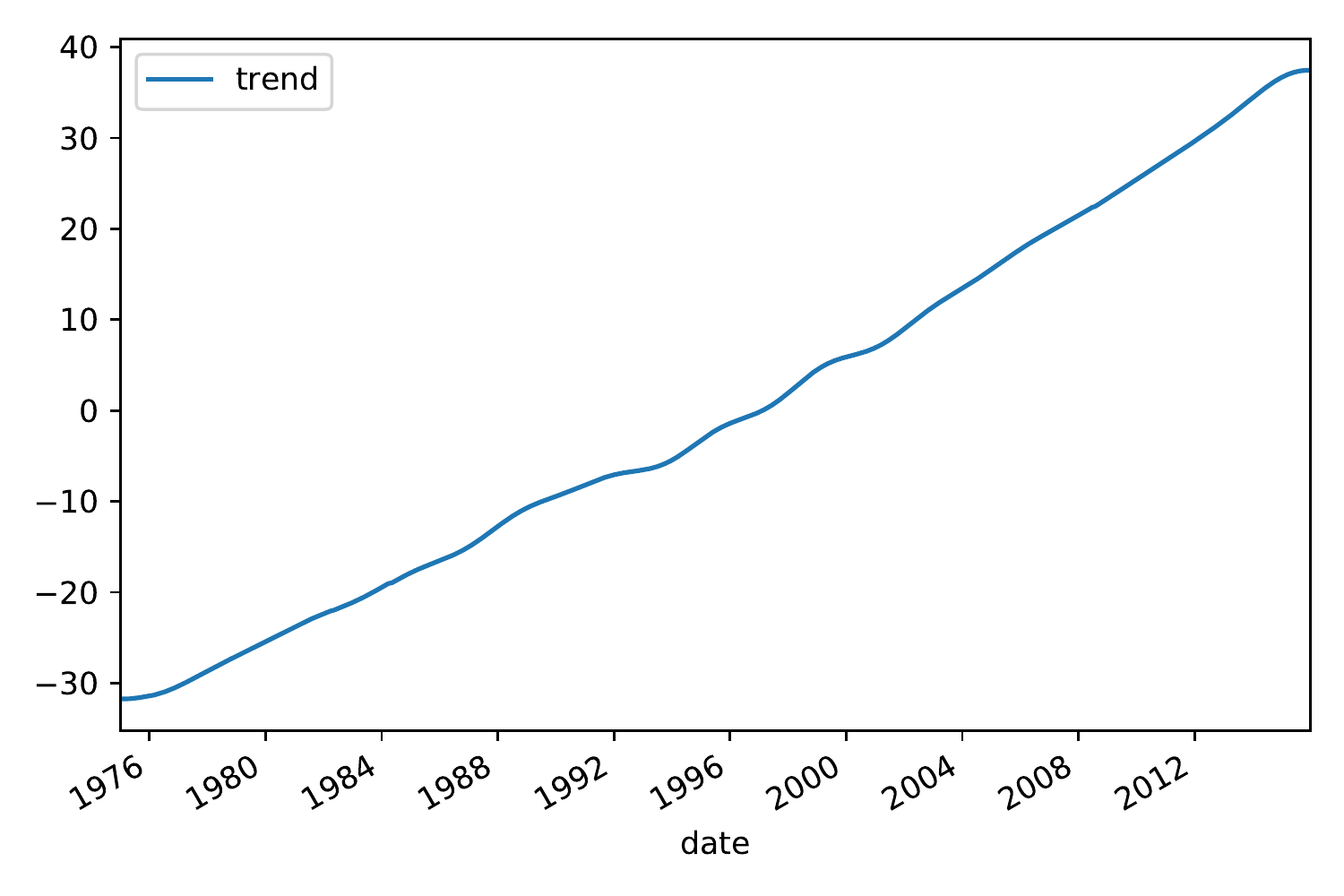}%
\caption{CO$_2$ time series, in parts per million (ppm), sampled weekly at the Mauna Loa station in the 1975-2015 period. A trend and a yearly oscillatory mode are clearly visible in the dataset. The right plot shows the obtained trend, an increase of approximately 70ppm is present during the analysed period.}
\label{fig:co2}
\end{figure}
%%%%%%%%%%%%%%%%%%%%%%
\section{Results}\label{results}
\begin{figure}
\centering
\includegraphics[scale=0.38]{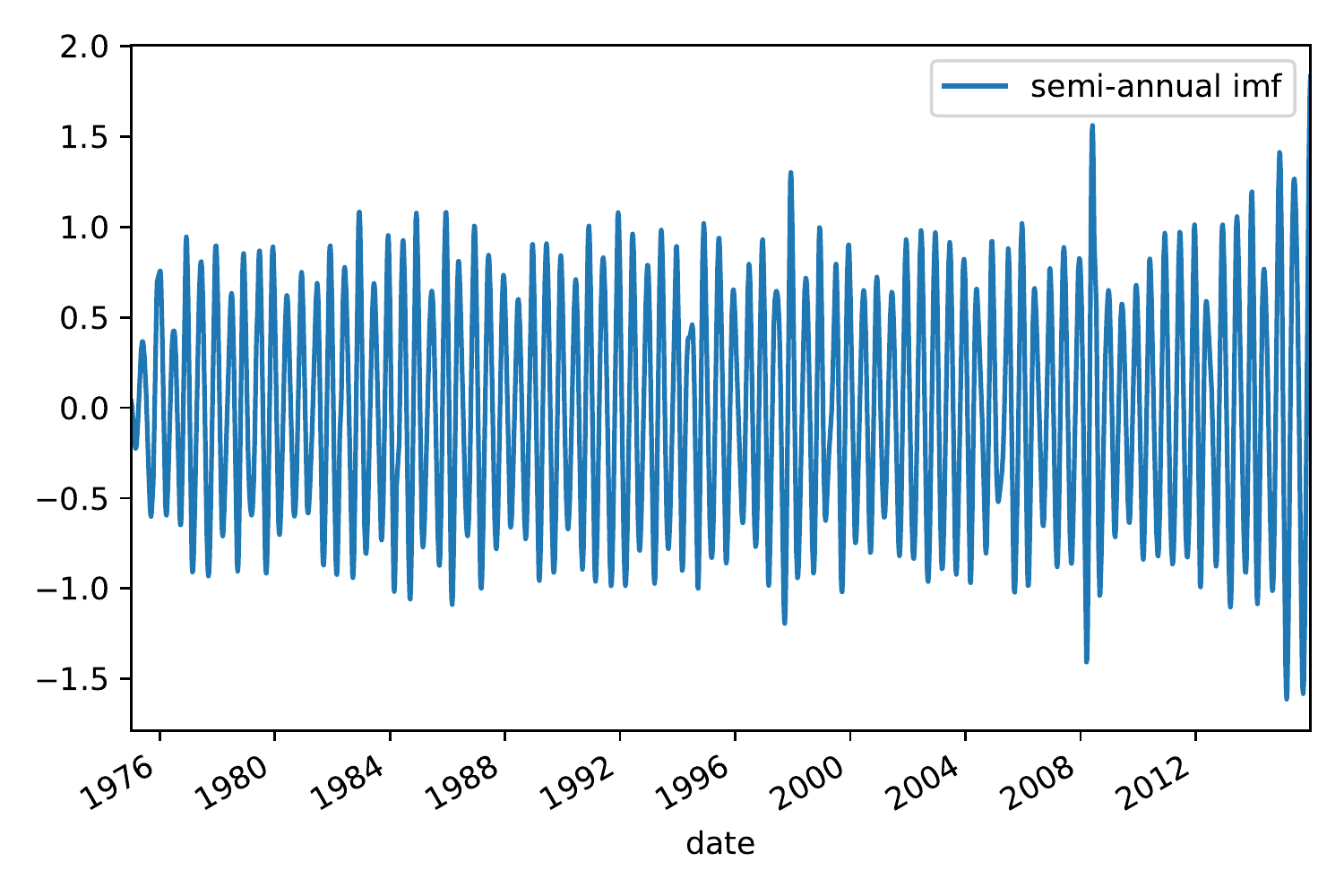}%
\qquad\qquad
\includegraphics[scale=0.38]{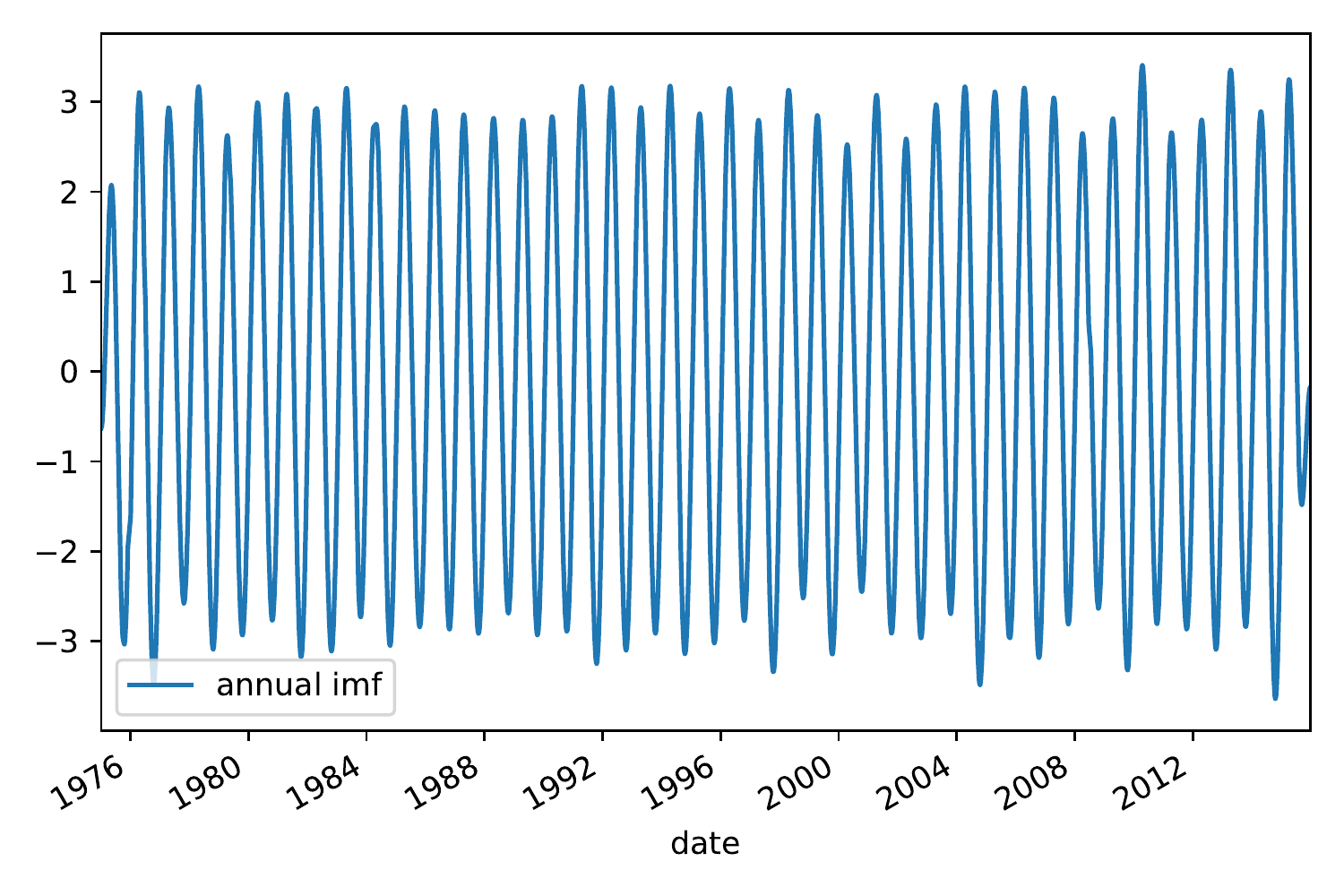}
\caption{Seasonal variability of the CO$_2$ data, as extracted by tvf-EMD algorithm: the semi-annual component is shown on the left while the annual component is on the right.}
\label{fig:semi}
\end{figure}

\begin{figure}
\centering
\includegraphics[scale=0.38]{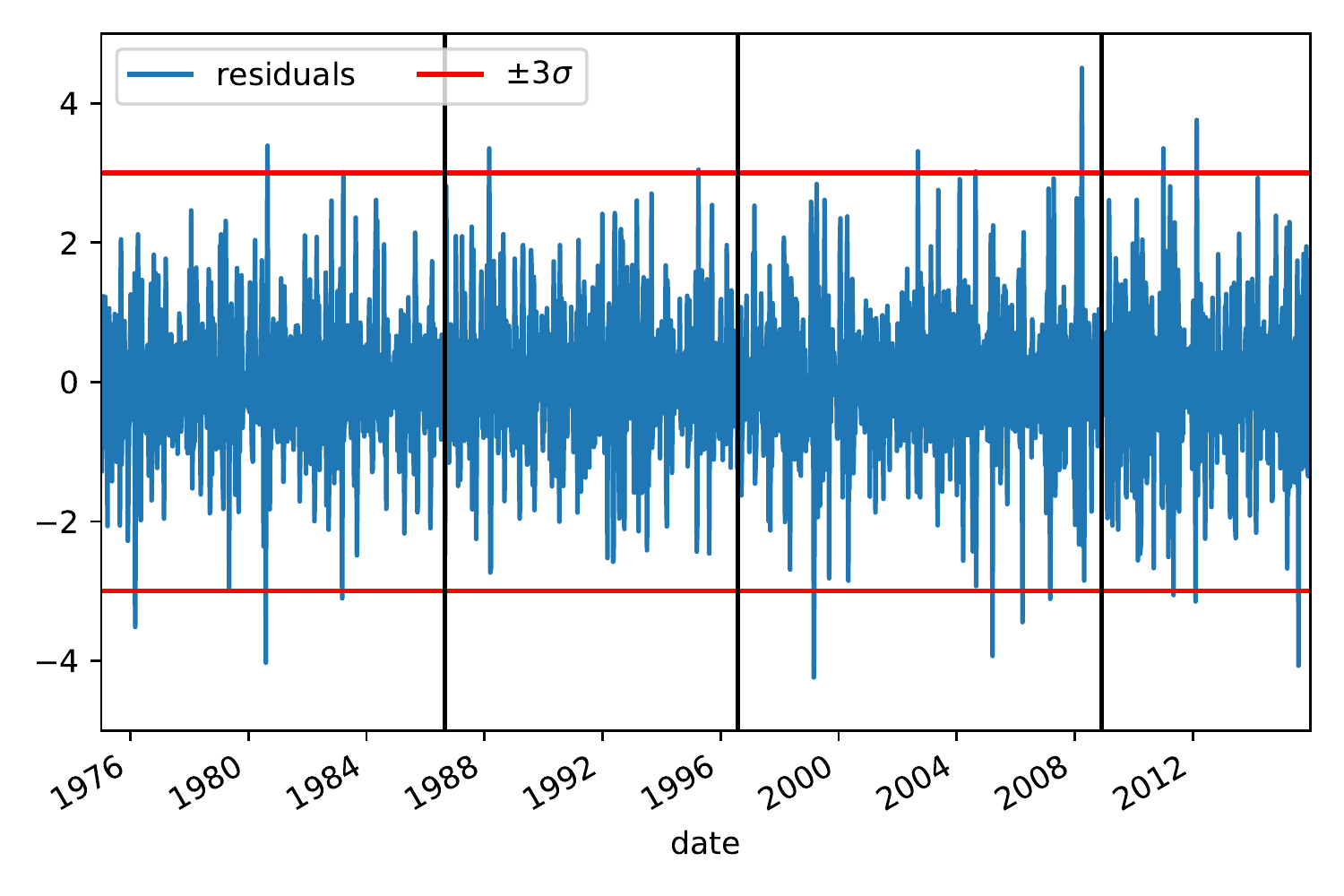}%
\qquad\qquad
\includegraphics[scale=0.38]{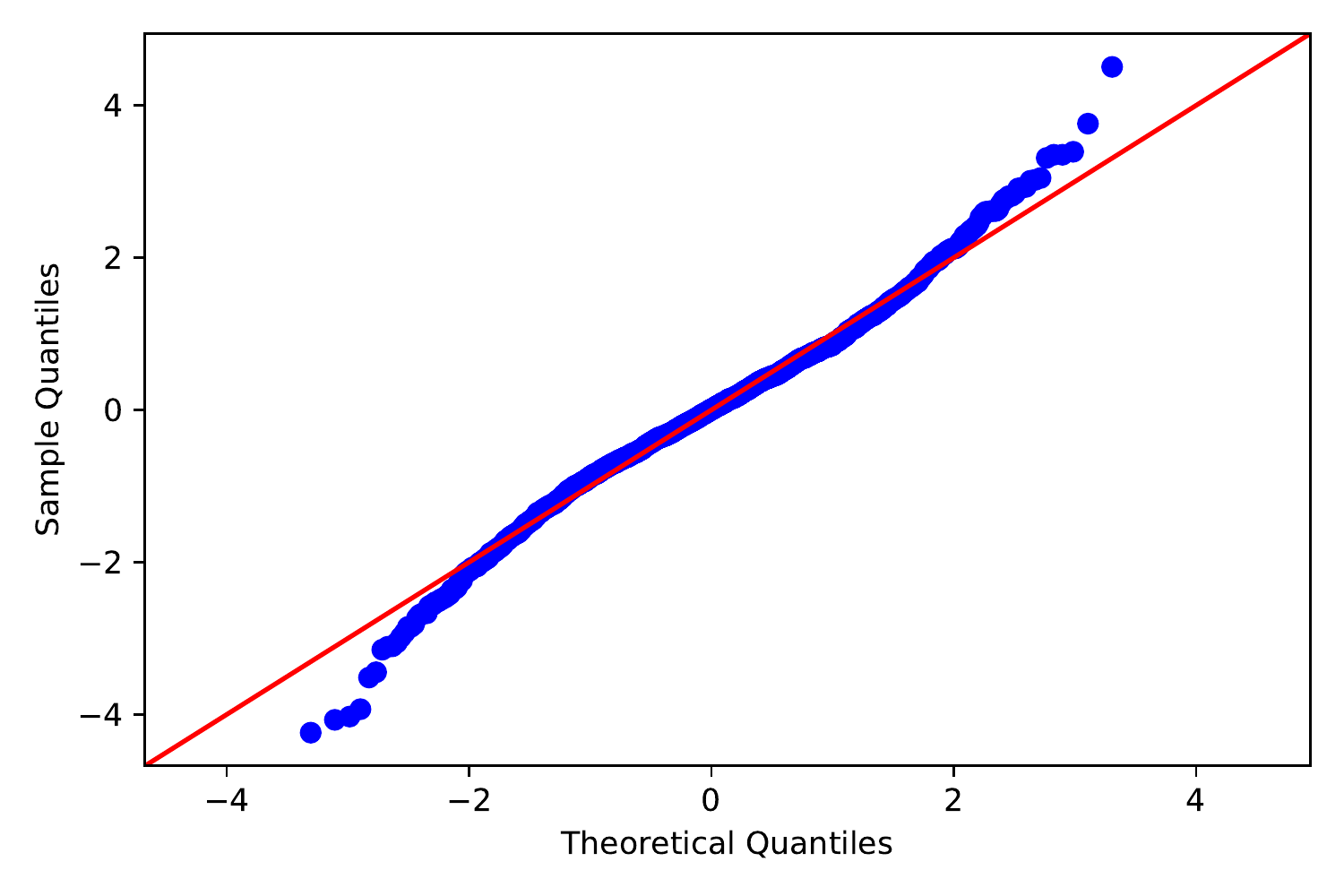}
\caption{Shown are the residuals, normalised to zero mean and unit standard deviation, as obtained after removal of the trend and of the yearly and seasonal oscillatory modes embedded in the CO$_2$ time series. The right plot shows instead the q-q plot for the residuals, which follow a normal distribution.}
\label{fig:res}
\end{figure}

Results of the analysis, obtained using the tvf-EMD algorithm described in \ref{methodology}, are summarised in this Section. In Figure \ref{fig:co2} the CO$_2$ time series data, in parts per million (ppm), sampled at the Mauna Loa station are reported. It can be seen that the variability in the data is due to both a trend term and seasonal oscillatory modes. The trend term obtained by the algorithm is also shown in Figure \ref{fig:co2}, from which it can be inferred that an increase of approximately 70 ppm in the average value of $CO_2$ occurs during the analysed period. The seasonal variability, as extracted by the tvf-EMD algorithm is reported in Figure \ref{fig:semi}, showing both the six month oscillatory mode and the yearly oscillatory mode. It can be seen that the amplitude of the yearly oscillatory mode is approximately three times larger than that of the semi-annual mode. The trend alone corresponds to 99.42\% of data variability, while trend and periodicities correspond to 99.98\% of the variability. Furthermore, both the yearly and semi-annual mode exhibit an amplitude modulation, which is more prominent for the semi-annual one. Finally, Figure \ref{fig:res} shows the residuals, i.e. the CO$_2$ data having subtracted the trend, the annual and the semi-annual components obtained and then normalized. Horizontal lines stand for $\pm 3 \sigma$, and values not between the two lines are considered outliers. To test if the residuals follow a normal distribution, Figure \ref{fig:res} also shows the q-q plot of the residuals, from which it can be seen that the residuals follows almost exactly a normal distribution. From the residual time series in Figure \ref{fig:res}, it can be seen that the occurrence of outliers is different over time. The plot has been divided into four regions, delimited by the vertical black lines, based on the times of occurrence of the last four solar cycles. It can be seen that in the first region there are 5 outliers, in the second one 2 while in the third one and in the fourth one there are 7 and 5 outliers, respectively. It is interesting to notice how periods with more outliers, i.e. the last two, correspond to solar cycles having smaller number of sunspots compared to the previous two cycles, but this needs further study. Furthermore, a more pronounced oscillation toward the end of the residual time series compared to the rest of the dataset is observed.

%%%%%%%%%%%%%%%%%%%%%%%%%%%%%%%%%%%%%%%%%%%%%%%%
\section{Discussion}
The adaptive algorithm tvf-EMD was employed to precisely separate and quantify the trend and seasonal components of the Mauna Loa $CO_2$ time series, allowing to investigate the amplitude modulation of the oscillatory modes. This algorithm was chosen for its ability to mitigate mode mixing, which occur when analysing noisy data due to the random distribution of extrema. The presence of noise can impact the envelope construction step of adaptive algorithms such as Empirical Mode Decomposition, leading to the extraction of mode mixed oscillatory modes. When mode mixing occur, oscillations of widely different scales are contained in the same oscillatory mode, which typically bear no physical meaning. tvf-EMD mitigates mode mixing using a stopping criterion based on the Loughlin instantaneous bandwidth. A threshold on the instantaneous bandwidth of the oscillatory modes to be extracted is used to stop the sifting process of the algorithm. Furthermore, B-splines are employed, and a frequency realignment step is also introduced. For the purpose of data analysis, we developed a Python version of the tvf-EMD algorithm (pytvfemd). Using an adaptive algorithm allowed to precisely extract the oscillatory modes in the data, which are both nonlinear and nonstationary. Having subtracted the trend and the seasonal variation, the residual time series was obtained and the time of occurrence of outliers could be investigated. It is found that the majority of the outliers occurs toward the end of the dataset, corresponding to solar cycles having smaller amplitudes, but this needs further investigation. A study of the IMFs of $CO$2, sunspot number and total solar irradiation time series, obtained using EMD, can be found in \cite{Barnhart_2011}. 

\section{Conclusions}\label{conclusions}
$CO_2$ data have been acquired since 1958 in the monitoring station located at Mauna Loa, Hawaii. Its isolated location and elevation makes it an ideal site for measuring this greenhouse gas far from anthropogenic sources and vegetation. The $CO_2$ time series analysed were acquired weekly during the 1975-2015 period. The data are characterised by an overall trend, which is due to anthropogenic emissions, and by a seasonal component made up of a six month and a yearly oscillation. Quantifying the variability of the $CO_2$ time series has relevant implications in modelling and forecasting, complementary to the investigation of sources and sinks of this greenhouse gas. Applying tvf-EMD allowed to obtain the seasonal variability of the $CO_2$ time series, in term of oscillatory modes not affected by mode mixing, and to investigate its amplitude modulation. Subtracting the seasonal variations the residuals were obtained, and their q-q plot revealed they follow a normal distribution. Finally, the time of occurrence of outliers in the residuals time series was investigated. It is found that a higher number of outliers occurs toward the end of the dataset, corresponding to solar cycle having a smaller amplitude. Further investigation is needed in this regard. A more pronounced oscillation of the residuals is also observed toward the end of the residuals time series.

%\clearpage
\section*{Acknowledgements}
Thanks are due to the National Oceanographic and Atmospheric Administration/Earth System Research Laboratory for providing
weekly $CO_2$ data at Mauna Loa. Data have been downloaded from \url{ftp://aftp.cmdl.noaa.gov/data/trace_gases/co2}.

\section*{}

\bibliography{bibl}
\bibliographystyle{ieeetr}

\end{document}